\documentclass[10pt,journal,compsoc]{IEEEtran}

\ifCLASSOPTIONcompsoc
  \usepackage[nocompress]{cite}
\else
  \usepackage{cite}
\fi

\usepackage{subfigure}
\usepackage{epsfig}
\usepackage{fancyhdr}
\usepackage{color}
\usepackage{multirow}
\usepackage{graphicx}
\usepackage{float}
\usepackage[numbers,sort&compress]{natbib}

\ifCLASSINFOpdf
\else
\fi
\begin{document}
%
\title{Benchmarking Big Data Systems: State-of-the-Art and Future Directions}
%
%
%
%

\author{
        Rui~Han, Zhen~Jia, Wanling~Gao, Xinhui~Tian, and Lei~Wang
\IEEEcompsocitemizethanks{
\IEEEcompsocthanksitem Rui~Han, Zhen~Jia, Wanling~Gao, Xinhui~Tian, and Lei~Wang are with the Institute of Computing Technology, Chinese Academy of Sciences, Beijing, China, 100190.
Email: \{hanrui,jiazhen,gaowanling,tianxinhui,wanglei2011\}@ict.ac.cn.\protect\\
}
}

\markboth{Technical Report. ICT, ACS}%
{Shell \MakeLowercase{\textit{et al.}}: Technical Report. ICT, ACS}

\IEEEtitleabstractindextext{%
\begin{abstract}
The great prosperity of big data systems such as Hadoop in recent years makes the benchmarking of these systems become crucial for both research and industry communities. The complexity, diversity, and rapid evolution of big data systems gives rise to various new challenges about how we design generators to produce data with the 4V properties (i.e. volume, velocity, variety and veracity), as well as implement application-specific but still comprehensive workloads. However, most of the existing big data benchmarks can be described as attempts to solve specific problems in benchmarking systems. This article investigates the state-of-the-art in benchmarking big data systems along with the future challenges to be addressed to realize a successful and efficient benchmark.
\end{abstract}

}

\maketitle

\IEEEdisplaynontitleabstractindextext

%
\IEEEpeerreviewmaketitle

\ifCLASSOPTIONcompsoc
\IEEEraisesectionheading{\section{Introduction}\label{sec:introduction}}
\else
\section{Introduction}
\label{sec:introduction}
\fi

%
%
%
%

\IEEEPARstart{B}{ig} data systems have gained unquestionable success in recent years and will continue its rapid development over the next decade. These systems cover many industrial and public service areas such as search engines, social networks, E-commerce sites and multimedia, as well as a variety of scientific research areas such as bioinformatics, environment, meteorology, and complex simulations of physics. Conceptually, big data are characterized by very large data volume and velocity, highly variety (diversity) in data types and sources, and stringent requirements of data veracity (fidelity). In the era of big data, the complexity, diversity of big data systems and the emergence of new systems driven by the exploration of big data values give rise to new challenges in how to benchmark and understand these systems successfully and efficiently. Enabling such benchmarking is essential so that system designers, programmers and researchers can optimize the performance and energy efficiency of big data systems and promote the development of big data technology.

Conceptually, a big data benchmark aims to generate \emph{application-specific} workloads and tests capable of processing \emph{big data} with the 4V properties (volume, velocity, variety and veracity) \cite{IBMplat2013} in order to produce meaningful evaluation results \cite{tay2011data}.
In this article, we survey the state-of-the-art of big data benchmarks in both academic and industry communities and provide a foundation towards building a successful big data benchmarks. The remainder of this paper is organized as follows. First of all, we present the methodology and requirements of benchmarking big data systems, which represents our insights into how to conduct successful benchmarking (Section \ref{Section: Big Data Benchmarking Methodology}). Next, the data generation techniques and the workload implementation techniques are presented to identifying the characteristics of big data benchmarks. Following this, a number of current benchmarks on evaluating big data systems are summarized (Section \ref{Section: State-of-the-art}). Finally, an open discussion of research challenges are presented to stimulate productive thinking, investigation, and development in this research area (Section \ref{Section: Research Challenges}) and the paper is summarized (Section \ref{Section: Conclusion}).






\section{Methodology and Requirements of Benchmarking Big Data Systems} \label{Section: Big Data Benchmarking Methodology}
In this section, we first present some basic concepts of big data systems (Section \ref{Section: Big Data Systems}), after which the methodology (Section \ref{Section: Benchmarking Methodology}) and requirements (Section \ref{Section: Benchmarking Requirements}) of benchmarking such systems are introduced.

\subsection{Big Data Systems}  \label{Section: Big Data Systems}
Technically, a big data system can be characterized by the \textbf{data} with the 4V properties and the \textbf{workloads} taking the data as inputs.
We first explain the 4V properties as follows.

\textbf{Volume.} Volume represents the amount/size of data such as TB, PB or ZB. Today, data are generated faster than ever. For example, about 2.5 quintillion bytes of data are created every day \cite{IBMplat2013} and this speed is expected to increase exponentially over the next decade according to International Data Corporation (IDC). In Facebook, there are 350 million photos updated and more than 500 TB data generated per day. Moreover, data volume has different meanings in workloads used to process different data sources. For example, in workloads for processing text data (e.g. \emph{Sort} \cite{sort2013} or \emph{WordCount}), the volume is represented by the amount of data (e.g. 1 TB or 1 PB text data). In social network workloads such as connected component, the volume is represented by the number of vertices (e.g. $2^{20}$ vertices) in social graphs.


\textbf{Velocity.} Velocity reflects the speed of generating, updating, or processing data. First of all, data velocity represents the \emph{data generation rate}, such as generating 10 TB per hour. Secondly, many big data applications such as e-commerce sites and social network  have continuously updating data. In this case, data velocity represents the \emph{data updating frequencies}. Finally, in streaming processing systems, data streams must be processed in real-time to keep up with their arriving speed. Hence data velocity represents the \emph{data processing speed}.


\textbf{Variety.} Variety denotes the range of data types and sources. The fast development of big data systems gives birth to a diversity of data types, which cover structured data (e.g. tables), unstructured data (e.g. text, images, audios, and videos), and semi-structured data (e.g. graph and web logs).


\textbf{Veracity.} Veracity reflects whether the data used in big data systems conform to the inherent and important characteristics of raw data. This property is important to guarantee the reality and credibility of benchmarking results.



Moreover, today's big data systems have a diversity of workloads whose behaviors (e.g. resource demands) are determined by two factors: (1) \textbf{Computation semantic}. This factor decides the implementation logic of workloads. For example, three Hadoop analytics workloads, Sort, WordCount, Grep, have different computation semantics (source codes), thus having different resource demands: Sort is an I/O-intensive workload, WordCount is a CPU-intensive workload with dominated integer calculations, and Grep has similar demands for CPU and I/O resources.
(2) \textbf{Software stacks}. A software stack consists of a set of programs working together to provide a fully functional solution. Model big data software stacks such as Hadoop \cite{Hadoop} and Spark \cite{Spark} usually provide rich libraries to facilitate the development of new applications. Hence a programmer can focus on writing a few lines of codes to implement an application (e.g. the Map and Reduce functions of a Hadoop MapReduce application) and leave parallelization, job scheduling, fault tolerance, and other tasks to the software stack.
Software stacks, therefore, cause considerable impact on a workload's behavior. For example, the Hadoop WordCount workload has similar behaviors to the Hadoop Bayes classification workload, but has different behaviors from the Spark WordCount workload \cite{jiacharacterizing}.


\subsection{Benchmarking Methodology}  \label{Section: Benchmarking Methodology}
Big data benchmarks are developed to evaluate and compare the performance of big data systems and architectures.
Successful and efficient benchmarking can provide realistic and accurate measuring of big data systems and thereby addressing two objectives. (1) Promoting the development of big data technology, i.e. developing new architectures (processors, memory systems, and network systems), innovative theories, algorithms and techniques to manage big data and extract their value and hidden knowledge. (2) Assisting system owners to make decisions for planning system features, tuning system configurations, validating deployment strategies, and conducting other efforts to improve systems performance. For example, benchmarking results can identify the performance bottlenecks in big data systems, thus guiding the optimization of system configuration and resource allocation.


Figure 1 shows a typical benchmarking methodology for big data systems and it consists of five stages. After selecting the application domain at stage 1, stage 2 surveys the representative applications in this domain. From these applications, this stage identifies data models from real data, data operations and workload patterns from real workloads, and evaluation metrics from performance and cost indicators. Based on the identification results, stage 3 implements data generation tools to produce data sets with the 4V properties and implements workloads to support application-specific benchmarking tests. Subsequently, stage 4 determines the target system, and prepares the input data and the benchmarking prescription used to test this system. A \emph{prescription} includes all the information needed to produce a benchmarking test, including input data, workloads, a method to generate test, and the evaluation metrics. Finally, the benchmark test is conducted at stage 5 and the evaluation result is analyzed and evaluated.



\begin{figure}[H]
\begin{center}
\includegraphics [width=0.43\textwidth]{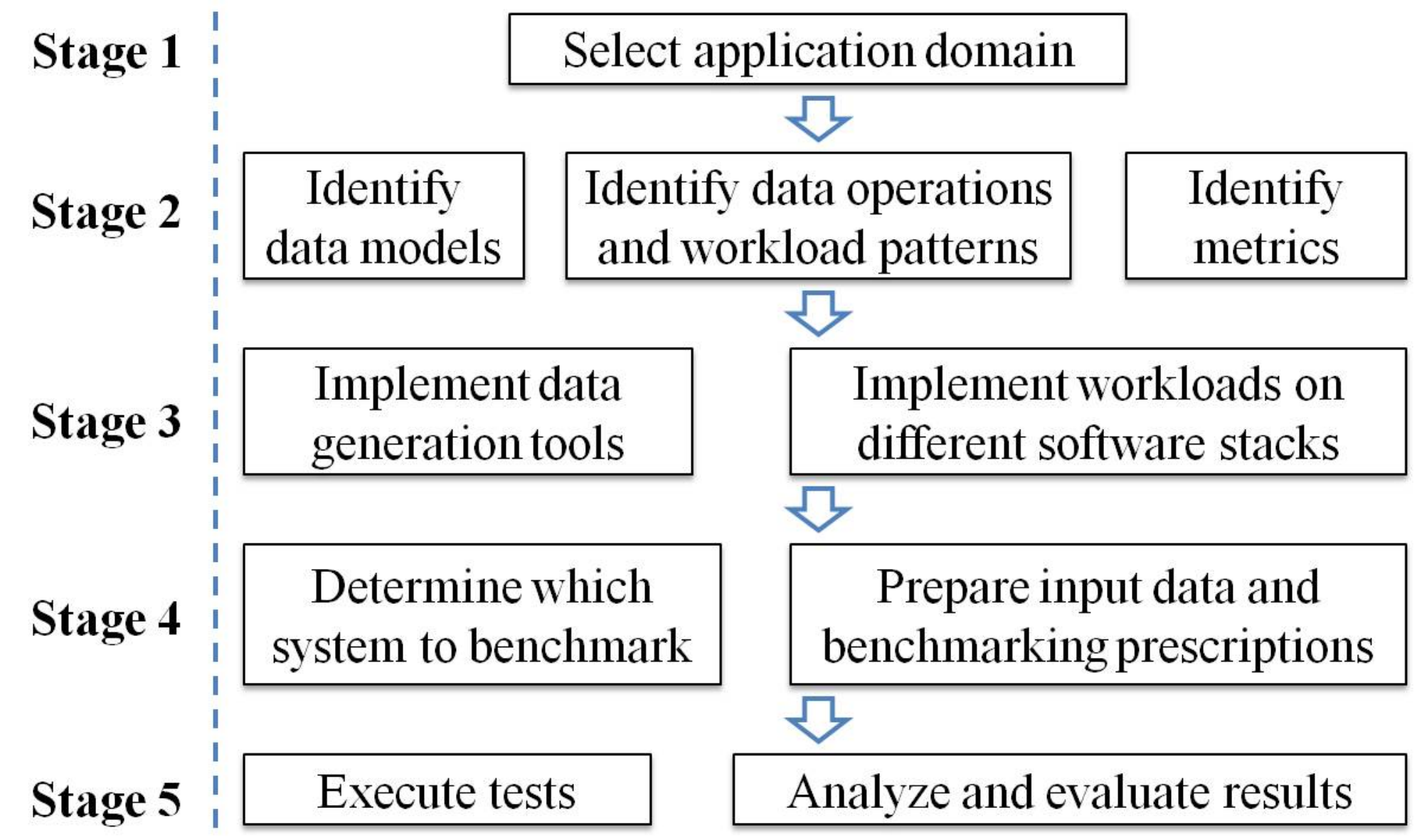}
\caption{The five-stage benchmarking methodology for big data systems}
\end{center}
\label{Fig: methodology}
\end{figure}

\subsection{Benchmarking Requirements}  \label{Section: Benchmarking Requirements}

In this section, we discuss the requirements of performing a fair, efficient and successful benchmarking test of big data systems.

\subsubsection{Generating Data with the 4V properties}

Some traditional benchmarks use real data as inputs of their workloads and thereby guarantee data veracity. However, the volume and velocity of real data cannot be flexibly adapted to different benchmarking requirements.
Based on our experience, we also noticed that in many practical scenarios, obtaining a variety of real data is difficult because many data owners are not willing to share their data due to confidential issues. In big data benchmarks, therefore, the consensus is to generate synthetic data as inputs of workloads on the basis of real data sets. Hence in synthetic data generation, preserving the 4V properties of big data is the foundation of producing meaningful and credible evaluation results. We discuss the data generation requirements based on the typical data generation processing shown in Figure 2.



At the first step, the data generation tools support the \textbf{variety} of big data by collecting real data to cover different data sources and types.
The tools can also generate synthetic data sets directly. This is because in practice such purely synthetic data can be used as inputs of some workloads such as the \emph{Sort} and \emph{WordCount} in Micro benchmarks, and the \emph{Read}, \emph{Write}, and \emph{Scan} belonging to basic database operations.

At the second step, each tool employs a data model to capture and preserve the important characteristics in one or multiple real data sets of a specific date type. For example, a text generator can apply Latent dirichlet allocation (LDA) \cite{blei2003latent} to describe the topic and word distributions in text data. This generator first learns from a real text data set to obtain a word dictionary. It then trains the parameters and a LDA model using this data set. Finally, it generates synthetic text data using the LDA model. To preserve data \textbf{veracity}, it is required that different models should be developed to capture the characteristic of real data of different types such as table, text, stream, and graph data.

At the third step, the \textbf{volume} and \textbf{velocity} can be controlled according to user requirements. For example, the data generation can be paralleled and distributed to multiple machines, thus supporting different data generation rates.

At the final step, after a data set is generated, the format conversion tools transform this data set into an appropriate format capable of being used as the input of a workload running on a specific system.



\begin{figure}[H]
\begin{center}
\includegraphics [width=0.5\textwidth]{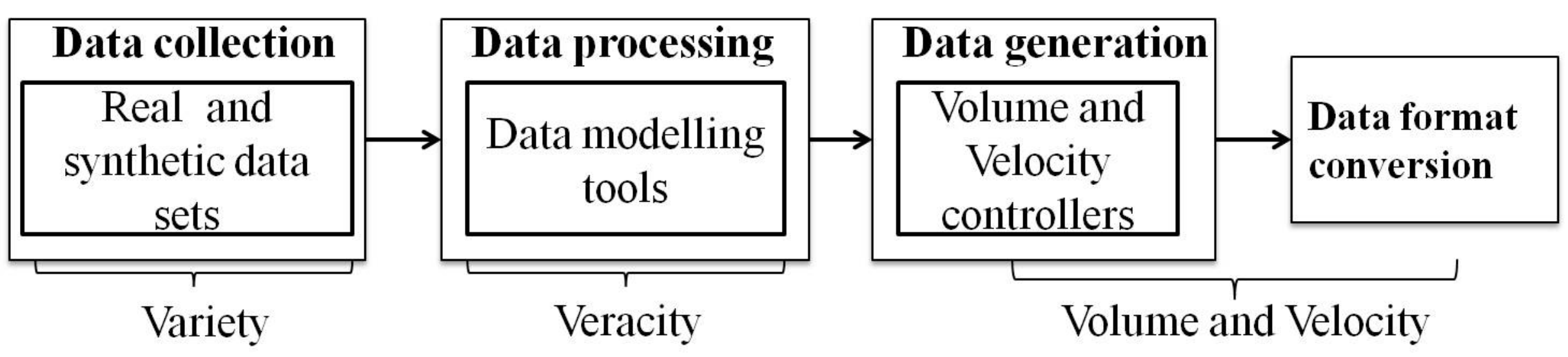}
\caption{The big data generation process}
\end{center}
\label{Fig: dataGenerationProcess}
\end{figure}


\subsubsection{Implementing Application-specific Workloads}
Margo Seltzer et al. pointed out that a benchmarking test is meaningful only when applying an application-specific workload \cite{tay2011data}. However, the diversity and rapid evaluation of big data systems means it is challenging to develop big data benchmarks to reflect various workload cases.
Hence in big data benchmarks, identifying the typical workload behaviors for an application domain is the prerequisite of implementing workloads to evaluate big data systems. We discuss the workload implementation requirements from the \emph{functional} perspective and the \emph{system} perspective, respectively.

\textbf{Functional perspective.} Given the complexity and diversity of workload behaviors in current big data systems, it is reasonable to say that no single set of behaviors is representative for all applications. Hence, it is necessary to abstract from the behaviors of different workloads to a general approach. This approach should identify representative workload behaviors in the application domain. In big data processing, the workload behaviors can be described as a set of operations and workload patterns.
\begin{itemize}
\item \emph{Operations} represent the abstracted processing actions (operators) on data sets. For example, \emph{select}, \emph{put}, \emph{get}, and \emph{delete} are the identified operations in database systems to operate table data.
\item \emph{Workload patterns} are designed to combine operations to form complex processing tasks. One identified workload pattern can contain one or multiple abstract operations as well as their workflow.  For example, a workload pattern representing a \emph{SQL query} can contain \emph{select} and \emph{put} operations, in which the \emph{select} operation executes first.
\end{itemize}

\textbf{System perspective.} The identified operations and patterns are designed to capture workloads' system-independent behaviors, i.e. the data processing operations and their sequences.
Based on abstracted operations and patterns, an abstracted workload can be constructed and this workload is independent of underlying systems.
From the system perspective, this abstract workload can be implemented on different software stacks (MPI, Hadoop and Spark) and thereby allows the comparison of systems of different type. For example, an abstract workload consisting of a sequence of \emph{read, write}, and \emph{update} operations can be used to compare a DBMS and the Hadoop MapReduce system.





\subsubsection{Benchmarkability Requirements}

The requirements of both data and workloads decide the worthiness of benchmarking results. In this section, we briefly discuss a list of benchmarkability requirements that decide the effectiveness of benchmarking big data systems.


\textbf{Usability.} Usability reflects users' experiences in using benchmarks and it is a combination of factors. In big data benchmarks, these factors include ease of deploying, configuring, and using across different software stacks; high benchmarking efficiencies; simple and understandable performance metrics; convenient user interfaces; and so on.


\textbf{Fair measurement.} A fair and sensible evaluation has twofold meanings. First, big data systems usually have many optional configurations, while these configurations have different combinations for optimal performance when the systems run in different hardware platforms. That is to say, using default configurations in measurement cannot guarantee fair measurement. Hence when comparing different big data systems in heterogeneous platforms, each system must be configured separately for fair comparison. For example, a big data system may have some specific configuration to improve its performance, this configuration therefore is not suitable for fair measurement. Second, repeatability is another important requirement of fail measurement. This requirement means the parameters of hardware and software configurations must be stately clearly so that the same result can be obtained when the evaluation is repeated several times. In particular, in cloud environment, there are multiple virtual machines (VMs) running in one physical machine (PM) and competing for compute resources. Hence we need to develop a comprehensive evaluation mechanism to effectively identify and estimate the impact of resource competition on benchmarking results, and try to avoid the uncertainties incurred by this resource competition.

\textbf{Measurability}. Metrics are crucial for quantifying, analyzing and evaluating benchmarking results. In big data systems, metrics (either single or multiple metrics) can typically be divided into two types: user-perceivable metrics and architecture metrics \cite{wang2014bigdatabench}. \emph{User-perceivable metrics} represent the metrics that matter for users; these metrics are usually observable and easy to be understood by users. Examples of user-perceivable metrics are the duration of a test, request latency, and throughput. While user-perceivable metrics are used to compare performances of workloads of the same category, \emph{architecture metrics} are designed to compare workloads from different categories. Examples of architecture metrics are million instructions per second (MIPS) and million floating-point operations per second (MFLOPS). In addition, these metrics should not only measure system performance, but also take energy consumption, cost efficiency into consideration.

\textbf{Extensibility.} The fast evolution of big data systems requires big data benchmarks not only keeping in pace with state-of-the-art techniques and underlying systems, but also taking their future changes into consideration. That is, big data benchmarks should be able to add new workloads or data sets with little or no change to the underlying algorithms and functions \cite{wang2014bigdatabench}.




\section{State-of-the-art} \label{Section: State-of-the-art}
In this section, we review related work on big data benchmarks from the perspectives of data generation and workload implementation.


\subsection{Data Generation Techniques} \label{Section: Data Generation Techniques}
We review data generation techniques in existing big data benchmarks according to the 4V properties of big data, as shown in Table 1.

\textbf{Volume.} Most of existing benchmarks generate scalable data as their workload inputs. By contrast, some benchmarks such as Hibench and LinkBench use both scalable and fixed-size data as inputs. Hence we call these benchmarks partially scalable in terms of data volume.

\textbf{Velocity.} To date, data velocity has not been adequately addressed in current benchmarks.
Some benchmarks such as LinkBench, CloudSuite and BigDataBench provide parallel strategies to enable the dynamic adjustment of data generation speed. However, another two equally important aspects of data velocity, namely the data updating and processing speeds, are not considered in these benchmarks. That is, these benchmarks are semi-controllable in terms of data velocity.
Other benchmarks are classified as un-controllable because they do not consider any aspect of data velocity.

\textbf{Variety.} Table 1 lists the data types and sources supported by each benchmark. These include structured data (tables), unstructured data (text, images videos and audios), and semi-structured data (graphs, web logs and resumes). We can observe that many benchmarks only support one type of data (e.g. the unstructured text data in Hibench or the structured table data in YCSB and TPC-DS). Some benchmark suites support multiple data types and BigDataBench supports the largest number of data sources.

\textbf{Veracity.} Veracity is one of the most challenging aspects in data generation.
In many benchmarks such as GridMix, SWIM, HiBench and YCSB, the generation process of synthetic data is independent of real raw data. For example, in HiBench \cite{huang2010hibench}, the synthetic data sets are either randomly generated using the programs in the Hadoop distribution or created using some statistic distributions. Data veracity therefore is un-considerable in these benchmarks.
By contrast, some benchmarks such as TPC-DS, LinkBench and BigDataBench capture and preserve the important characteristics of real data by identifying data models. The synthetic data are then generated using the constructed data model, thus partially considering data veracity.
For example, TPC-DS \cite{ghazal2013bigbench} implements a multi-dimensional data generator (MUDD). MUDD generates most of data using traditional synthetic distributions such as a Gaussian distribution. On the other hand, MUDD generates a small portion of crucial data sets using more realistic distributions derived from real data. Moreover, in BigDataBench \cite{wang2014bigdatabench}, different data models are employed to capture and preserve the important characteristics of raw data of different types (e.g. table, text, and table).


\begin{table*} [!t]
\caption{Comparison of data generation techniques in existing big data benchmarks.}
\centering
\begin{tabular}{|c|c|c|c|c|c|}
\hline
\textbf{Benchmark efforts} & \textbf{Data Volume} & \textbf{Data Velocity} & \multicolumn{2}{c|}{\textbf{Data Variety}}   & \textbf{Data Veracity}\\
\cline{4-5}
&&& Data types & Data sources &\\
\hline
Sort~\cite{sort} & Scalable & Un-controllable & Unstructured data& Texts&Un-considered\\
\hline
DFSIO~\cite{DFSIO} & Scalable & Un-controllable & Unstructured data& Texts&Un-considered\\
\hline
MRBench~\cite{kim2008mrbench} & Scalable & Un-controllable & Structured data& Tables &Un-considered\\
\hline
GridMix \cite{GridMix2013} & Scalable & Un-controllable & Unstructured data& Texts&Un-considered\\
\hline
PigMix \cite{PigMix2013} & Scalable & Un-controllable & Unstructured data& Texts& Un-considered\\
\hline
SWIM~\cite{chen2012interactive} & Scalable & Un-controllable & Unstructured data& Texts& Un-considered\\
\hline
Hibench \cite{huang2010hibench} & Partially scalable &Un-controllable & Unstructured data& Texts &Un-considered\\
\hline
CALDA \cite{pavlo2009comparison} & Scalable & Un-controllable &Structured& Tables, texts & Un-considered\\
 & & & and unstructured data& & \\
\hline
AMPLab & Scalable & Un-controllable &Structured and& Tables, texts & Un-considered\\
benchmark \cite{AMPBenchmark} & & &unstructured data& & \\
\hline
BigBench \cite{ghazal2013bigbench} &Scalable & Semi-controllable &Structured, semi-structured& Tables, web logs& Partially Considered\\
 &  & & and unstructured data& and texts  & \\
\hline
LinkBench \cite{armstrong2013linkbench} &Partially scalable & Semi-controllable &Semi-structured data& Graphs & Partially Considered\\
\hline
TPC-DS \cite{tpcds} &Scalable & Semi-controllable &Structured data& Tables & Partially Considered\\
\hline
BG benchmark \cite{barahmand2013bg} &Scalable & Semi-controllable &Structured data& Tables & Un-considered\\
\hline
YCSB~\cite{cooper2010benchmarking} & Scalable & Un-controllable &Structured data& Tables & Un-considered\\
\hline
CloudSuite \cite{ferdman2011clearing} &Partially scalable & Semi-controllable &Structured, semi-structured&Tables, resumes,& Partially Considered\\
 &  & & and unstructured data& graphs and texts  & \\
\hline
BigDataBench \cite{wang2014bigdatabench} &Scalable & Semi-controllable &Structured, semi-structured&Tables, resumes,&Partially considered\\
 &  & & and unstructured data& graphs, texts, images& \\
 &  & & &videos and audios  & \\
\hline
\end{tabular}
\label{tab:applications}
\end{table*}

\subsection{Workload Implementation Techniques}
In the context of benchmarking big data systems, many existing big data benchmarks implement workloads to evaluate specific types of systems or architectures. As listed in Table 2, Sort, DFSIO, MRBenchmark, GridMix, PigMix and SWIM are designed for Hadoop systems, LinkBench and TPC-DS implement workloads for testing DBMSs. HiBench, CALDA and BigBench are developed to compare the performance between Hadoop and Hives (or DBMSs).
Specifically, CALDA compare two parallel SQL DBMSs (i.e. DBMS-X and Vertica) with Hadoop \cite{pavlo2009comparison}. The same workloads are used to test four typical SQL driven systems, including one database (Redshift), one data warehousing systems (Hive), and two engines (Spark and Impala) \cite{AMPBenchmark}. TPC-DS is TPC's latest decision support benchmark \cite{TPC2013} designed to test the performance of DBMSs in decision support systems. Developed from TPC-DS by adding a web log generator and a review generator, BigBench aims to compare the Teradata Aster DBMS and Hadoop \cite{ghazal2013bigbench}.


Some other benchmarks target at evaluating NoSQL databases or architectures. Yahoo! Cloud Serving Benchmark (YCSB) benchmark compares two non-relational databases (Cassandra and HBase) against one geographically distributed database (PNUTS) and a traditional relational database (MySQL) \cite{cooper2010benchmarking}. The CloudSuite benchmark \cite{ferdman2011clearing} is implemented to test cloud service architectures. 
To the best of our knowledge, BigDataBench is the only big data benchmark that implements a comprehensive suite of workloads covering micro benchmarks, three dominant internet services (i.e. Search Engine, Social Network, and E-commerce), and two fast emerging big data domains (multimedia and bioinformatics).




From the perspective of benchmarking users, Table 2 roughly divides the current big data workloads into two types. (1) \textbf{Online services}: these workloads are sensitive to the response delay, i.e. the time interval between the arrival and departure moments of a service request. Examples of workloads belonging to this category are OLTP and web search queries. (2) \textbf{Offline analytics}: these workloads usually perform complex and time-consuming computations on big data. Examples of workloads for testing offline services are machine learning algorithms such as k-means clustering and Naive Bayes classification.


\begin{table*}[!t]
\caption{Comparison of implemented workloads and supported software stacks in existing big data benchmarks.}
\centering
\newcommand{\tabincell}[2]{\begin{tabular}{@{}#1@{}}#2\end{tabular}}
\begin{tabular}{|c|c|l|c|}
\hline
\multirow{2}*{\textbf{Benchmark efforts}} & \multicolumn{2}{|c|}{\textbf{Workloads}} & \multirow{2}*{\textbf{Software stacks}}\\
\cline{2-3}
& Type & Operations &\\
\hline
Sort~\cite{sort} & \tabincell{c}{Offline analytics} &\tabincell{l}{Sort} & Hadoop\\
\hline
DFSIO~\cite{DFSIO} & \tabincell{c}{Offline analytics} &\tabincell{l}{Generate, read, write, append,\\and remove data for MapReduce jobs} & Hadoop\\
\hline
MRBench \cite{kim2008mrbench} & \tabincell{c}{Online services} & \tabincell{l}{MapReduce jobs transformed from 22 \\TPC-H queries} & Hadoop\\
\hline
GridMix \cite{GridMix2013} & \tabincell{c}{Online services} & Sort, sampling a large dataset & Hadoop\\
\hline
PigMix \cite{PigMix2013} & \tabincell{c}{Online services} & 12 data queries & Hadoop\\
\hline
SWIM~\cite{chen2012interactive} & \tabincell{c}{Offline analytics} &\tabincell{l}{Synthetic MapReduce jobs of reading, writing, \\shuffling and sorting data} & Hadoop\\
\hline
\multirow{2}*{HiBench \cite{huang2010hibench}} & \tabincell{c}{Offline analytics} & \tabincell{l}{Sort, WordCount, TeraSort, PageRank,\\ K-means, Bayes classification, Index}
& \multirow{2}*{Hadoop and Hive}\\
\hline
\tabincell{c}{CALDA} \cite{pavlo2009comparison} & \tabincell{c}{Online services} & \tabincell{l}{Load, scan, select, aggregate \\ and join data, count URL links }&Hadoop and DBMSs\\
\hline
{AMPLab benchmark \cite{AMPBenchmark}} &\tabincell{c}{Online services} & \tabincell{l}{Part of CALDA workloads (scan, aggregate \\ and join) and PageRank} & \tabincell{c}{Redshift, Hive, Shark, \\ Impala and Tez}\\
\hline
\multirow{2}*{BigBench \cite{ghazal2013bigbench}} &\tabincell{c}{Online services} & \tabincell{l}{Database operations \\ (select, create and drop tables)} & \multirow{2}*{Hadoop and DBMSs}\\
\cline{2-3}
& \tabincell{c}{Offline analytics} & K-means, classification &\\
\hline
LinkBench \cite{armstrong2013linkbench} &\tabincell{c}{Online services} &\tabincell{l}{Database operations such as select, insert,\\ update,and delete;\\association range queries and count queries} & DBMS\\
\hline
TPC-DS \cite{tpcds} &\tabincell{c}{Online services} & Data loading, queries and maintenance & DBMS\\
\hline
BG benchmark \cite{barahmand2013bg} &\tabincell{c}{Online services} & Reading and updating databases & DBMS and NoSQL systems\\
\hline
YCSB \cite{cooper2010benchmarking} & \tabincell{c}{Online services} & OLTP (read, write, scan, update) & NoSQL systems\\
\hline
\multirow{2}*{CloudSuite \cite{ferdman2011clearing}} &\tabincell{c}{Online services} & YCSB workloads & \multirow{2}*{\tabincell{c}{NoSQL systems,\\ Hadoop, GraphLab}}\\
\cline{2-3}
& \tabincell{c}{Offline analytics} & Text classification, WordCount &\\
\hline
\multirow{3}*{BigDataBench \cite{wang2014bigdatabench}} &\tabincell{c}{Online services} & Database operations (read, write, scan) &
\multirow{3}*{\tabincell{c}{Hadoop, DBMSs, NoSQL systems,\\ Hive, Impala, Hbase, MPI, \\ Shark, Libc, and other \\ real-time analytics systems}}\\
\cline{2-3}
& \tabincell{c}{Offline analytics} &
\tabincell{l}
{\emph{1. Micro Benchmarks} (sort, grep, WordCount, CFS);
\\
\emph{2. Search engine workloads} (index, PageRank);
\\
\emph{3. Social network workloads} (connected \\ components (CC), K-means and BFS);
\\
\tabincell{l}{\emph{4. E-commerce site workloads} (Relational database\\ queries (select, aggregate and join), \\ collaborative filtering (CF) and Naive Bayes;}
\\
\emph{5. Multimedia analytics workloads} (BasicMPEG,\\ SIFT, DBN, Speech Recognition, Ray Tracing, \\ Image Segmentation, Face Detection);
\\
\emph{6. Bioinformatics workloads} (SAND and BLAST)
} &\\
\hline
\end{tabular}
\label{tab:applications}
\end{table*} 
\section{Research Challenges} \label{Section: Research Challenges}




Most of the existing works related to benchmarking big data systems can be viewed as attempts to solve specific benchmarking problems. To date, many aspects of benchmark big data systems remain unexplored.
Considering the emergence of new big data applications and the rapid evolution of big data systems, we believe an incremental and iterative approach is necessary to conduct the investigations on big data benchmarks. In this section, we summarize some of the research challenges to be addressed for a successful benchmarking.


\subsection{Generating Data with Velocity and Veracity Properties}

One fundamental challenging of successfully benchmarking big data systems is about generating data with the 4V properties. Given that data volume and variety have been well supported in today's benchmarks, how to generate data with velocity and veracity properties have not yet adequately solved.


\textbf{Controllable data velocity.} This controllability has two meanings. First, existing big data benchmarks only consider different data generation speeds. Different data updating frequencies and processing speeds should be reflected in future big data generators. Second, current data velocity is supported using parallel strategies; that is, data velocity can be controlled by deploying different numbers of parallel data generators. To support the control mechanism at a finer level of granularity, future generators can control data velocity by adjusting the efficiency of the data generation algorithms themselves.
For example, in a graph data generator running on Spark (an in-memory computing platform), the data generation speed can be controlled by adjusting the allocated memory resources.

\textbf{Metrics to evaluate data veracity.} As discussed in Section \ref{Section: Data Generation Techniques}, applying data models to capture and preserve important characteristics of real data is an efficient way to keep data veracity in synthetic data generation.
However, how to measure the conformity of the generated synthetic data to the raw data is still an open question; that is, metrics need to be developed to evaluate data veracity, thus guiding the improvement and optimization of the constructed models. Two types of evaluation metrics can be developed: (1) metrics to compare the raw data and the constructed data models; (2) metrics to compare the raw data and the synthetic data.

This problem is compounded when considering different data types and sources. For example, to compare real text data set and synthetic data, we first need to derive the topic and word distributions from these data sets. Next, statistical metrics such as Kullback-Leibler divergence can be applied to compare the similarity between two distributions. Similarly, when considering table, graph or even stream data, some other metrics should be developed.



 
\subsection{Identifying Dwarf Big Data Workloads} 

The fast development of big data systems has lead to a number of successful application domains such as scientific analytics, search engines, social networks, and streaming process. Each of these application domains is the focus of one or multiple big data platform efforts. A successful big data benchmark, therefore, should provide workloads with representativeness and wide coverage. However, the complexity and diversity of big data systems impose great challenges on workload selection, as it is unpractical to implement all big data workloads. Within this context, identifying a minimal set of dwarf workloads to represent diverse big data workloads provides an effective approach ~\cite{asanovic2006landscape}. These dwarf workloads have two appealing features. First, they represent the highly abstractions of computation (data operations), communication and workload patterns frequently appearing in big data processing. Second, they compose a minimum set of necessary functionality~\cite{minimumSet}, thus provide guidelines for performance optimization~\cite{asanovic2006landscape}.
 


Considering the diversity of big data domains and the heterogeneity of big data systems, we summarize the challenges in identifying dwarf workloads as follows. (1) As big data systems being applied in more and more domains, it is difficult to cover all these domains. (2) Even in some popular domains such as big data analytics, identifying the frequently appearing operations from a wide range of data mining and machine learning algorithms is not trivial. (3) The problem is compounded when considering the variety of data in big data systems, in which 80\% of operations are conducted on unstructured data \cite{multimediagrowth}.



\subsection{Automating Test Generation}

In practice, generating benchmarking tests from identified dwarf workloads including data operations and workload patterns may be beyond the capabilities of the average benchmark users. Hence going mainstream with this framework requires the development of an environment that supports benchmark users in profiling the behaviors of applications in the target domain, as well as offers a repository of reusable data models, operations and workload patterns to simplify the generation of input data, workloads and tests running on state-of-the-art software stacks.

Moreover, the automatic generation of benchmarking tests also needs a parameterized framework that enables the flexible adjustment in input data and workload generation, thus meeting the requirements of different benchmarking scenarios. Developing such a framework requires addressing two challenges: (1) based on the identified data models and workload operations, the framework should derive optional parameters that allow users use different parameter configurations to customize their input data and workload generation. (2) The framework should study and quantify the corresponding relationship between input data and workloads, thus imposing constraints on the provided parameters to guarantee the reasonable matching in input data and workload generation.

\subsection{Characterizing Big Data Workloads}
 
Understanding and characterizing workloads provides the foundation for optimizing and promoting big data systems. A typical example can be the system architecture design driven by advances in processor designs, in which significant efforts are contributed to workload characterization in order to obtain the optimization guides or implications of the next generation processors~\cite{reddi2010web,jiacharacterization,ferdman2011clearing,wang2014bigdatabench}. Within the context of big data systems, there are three major challenges in big data workload characterization.

\textbf{Obtaining performance data}. In traditional workloads such as Parsec~\cite{bienia2008parsec}, architecture simulators are typically used to acquire fine grain performance data in order to characterize these workloads from the perspective of micro-architecture. Such simulation-based workload characterization is prohibitively expensive for big data workloads. This is because data big data workloads always process data with deep software stacks, which means it may need thousands of billions of instructions to complete one workload and this processing can take several months when running on simulators. Hardware performance counters offer an cost-efficient solution to obtain processors' predefined micro-architecture level statistics such as cache misses and number of retired instructions. However, they can only provide coarse-grained architecture dependant data, but lack the support of fine-grain performance data such as cache false sharing. In addition, Dimitrov et al.~\cite{dimitrov2013memory} can get the fine grain data by the analysis the memory DIMM traces, but this data acquisition replies on special hardware. Moreover, when characterizations big data workloads from other perspectives such as memory and storage subsystem, there exists similar challenges of obtaining fine-grained performance data without simulation or special hardware support, which prohibits the thorough understanding of workload behaviors.

\textbf{Determining a suitable data volume}. In big data workloads, the acquired performance data vary with data volumes~\cite{gaoisca,jia2014implications}, and thus data volumes affect the result of workloads characterization.
In big data systems, data volumes depend on a list of factors including the scale of cluster, the configuration of each node, and the tested workloads. Hence a range of optional data volumes exist when testing a workload on a specific cluster. How to determine a suitable data volume is still an open question.

\textbf{Subsetting big data workloads}. Today's big data systems have a large number of workloads and these workload are changing frequently, which is called workload churns~\cite{barroso2013datacenter}. This means even on the real hardware (rather than using simulators), the evaluation can be very time consuming. It is therefore essential to develop effective approaches to remove redundant workloads and generate a subset of workloads with manageable number, while still keeping their representativeness of the whole workload set. Jia et al. propose an approach that derives a subset of representative workloads from the perspective of micro-architecture~\cite{jiasub}. How to subset workloads from other perspectives still needs to be investigated. Furthermore, how to adapt to the workload churns (that is, new workloads are continuously added) in workload subsetting is another challenge to be addressed.
 


    


\subsection{Generating Realistic Mixed Workloads}

With the fast development of big data systems and applications, a diverse mix of workloads share a common computing infrastructure in modern cloud data centers. These workloads have different system behaviours and input data sizes (e.g. ranging from KB to PB), and their behaviors also heavily rely on the underlying software stacks such as Hadoop, Spark, Hive and Impala. Moreover, their dynamic arrival patterns including request/job arriving rates and sequences are equally important aspects of workload features to be considered. As big data systems such as Hadoop mature, the pressure to benchmark and understand these mixed workloads rises.

To generate realistic mixed workloads such that trustworthy benchmarking reflecting the practical data center scenarios can be conducted, three major challenges should be addressed. First, it is difficult to just use synthetic workloads to emulate the behaviors of workloads with highly diverse features in terms of workload types, input sizes and software sizes. Hence it is necessary to generate benchmark results using actual workloads. Second, we believe profiling history logs of real applications, namely actual workload traces, is a good way to obtain realistic arrival patterns. How to use this profiling information to guide the generation of actual workloads is still an open question. Finally, it is challenging to produce workloads at different scales to meet the requirements of different benchmarking scenarios, while still keep the realistic mix of big data workloads.
To the best of our knowledge, none of exiting big data benchmarks have solved all the above challenges.

\subsection{Benchmarking Emerging In-memory Computing Systems}
As the requirement of low latency computation increases, many researchers now pay more attention on how to use memory more efficiently to improve performance of data processing. A typical example of applying this in-memory computing paradigm is Spark, a big data platform that adopts memory locality and intermediate result caching to speed up big data processing. At present, the Spark community has built an ecosystem to support various application domains including machine learning, SQL query, graph computation, streaming applications, and R language.


The features of in-memory computing systems give rise to two major challenges in benchmarking them. First, memory has a larger impact on system performance than other resources, hence the workloads' efficiently of using memory resources significantly impact benchmarking results. However, this memory usage is difficult to control when benchmarking some in-memory computing systems. For example, Spark relies on JVM for memory management and this makes it difficult to monitor the actual memory usage in benchmarking. Second, in-memory computing systems such as Spark apply compression and serialization techniques to reduce the memory usage by significantly increasing the CPU utilization. In this case, how to measure the CPU usage and other low level performance metrics is another challenging to be addressed.
 
\subsection{Supporting Heterogeneous Hardware Platforms}

With the fast development of technology, the emerged hardware platforms significantly change the way about how to process data and show a promising prospect to improve data processing efficiency. For example, the heterogeneous platforms of Xeon+General-purpose computing on graphics processing units (GPGPU) and Xeon+Many Integrated Core (MIC) can significantly improve the processing speed of HPC applications. However, to date, both platforms are only limited to the HPC area; that is, the diversity of big data applications are not fully considered in these platforms. For such an issue, big data benchmarks should be developed to evaluate and compare workloads in state-of-the-practice heterogeneous platforms. The evaluation results are expected to show: (1) whether any platform can consistently win in terms of both performance and energy efficiency for all big data applications, and (2) for each class of big data applications, we hope to find some specific platform that can realize better performance and energy efficiency for them. To support the evaluation of an application, current big data benchmarks should be extended to provide a uniform interface to enable this application running in different platforms. In order to perform apples-to-apples comparisons, this application should also be running in the same software stack.

\section{Conclusion} \label{Section: Conclusion}

With the rapid development of information technology, big data systems have emerged to manage and process data with high requirements of volume, velocity, variety and veracity. These emerging systems have given rise to various new challenges about how to develop a new generation of benchmarks. In this paper, we summarize the lessons we have learned and challenges in developing big data benchmarks mainly from two aspects: (1) how to develop data generators capable of preserving the 4V properties of big data; (2) how to implement application-specific workloads while still covering  a diversity of typical application scenarios and supporting different system implementations and software stacks. Following these two aspects, we review existing benchmarking techniques on big data systems and present some future research directions.
The work presented in this paper represents our effort towards building a truly representative, comprehensive and cost-efficient big data benchmark and we encourage more investigations and developments to make this become a reality. 

\section*{Acknowledgments}
This technical report is a significant extended version of its preliminary version entitled "On Big Data Benchmarking", which is published in BPOE-4 (Co-located with ASPLOS 2014)~\cite{han2014big}.



\ifCLASSOPTIONcaptionsoff
  \newpage
\fi

\bibliographystyle{IEEEtran}
\bibliography{reference}

\end{document}